\pdfoutput=1

\documentclass[11pt]{article}

\usepackage{acl}

\usepackage{times}
\usepackage{latexsym}
\usepackage{multirow}
\usepackage{graphicx}
\usepackage{amsmath}
\usepackage{makecell}
\usepackage{amssymb}
\usepackage{float}
\usepackage{enumitem}
\usepackage{hyperref}

\usepackage[T1]{fontenc}

\usepackage[utf8]{inputenc}

\usepackage{microtype}

\title{ReACC: A Retrieval-Augmented Code Completion Framework}

\author{
Shuai Lu$^{1}$, 
Nan Duan$^{1}$, 
Hojae Han$^{2}\thanks{\ \ Work done during internship at Microsoft.}$\hspace{0.3em}, 
Daya Guo$^{3*}$,\\
{\bf Seung-won Hwang$^{2}$,
Alexey Svyatkovskiy$^{4}$}\hspace{0.5em}\\
$^{1}$Microsoft Research Asia \quad
$^{2}$Seoul National University \\
$^{3}$Sun Yat-sen University \quad
$^{4}$Microsoft Devdiv \\
{\tt \{shuailu,nanduan,alsvyatk\}@microsoft.com} \\
{\tt \{stovecat,seungwonh\}@snu.ac.kr} \\
{\tt guody5@mail2.sysu.edu.cn}
}

\begin{document}
\maketitle

\begin{abstract}
Code completion, which aims to predict the following code token(s) according to the code context, can improve the productivity of software development. 
Recent work has proved that statistical language modeling with transformers can greatly improve the performance in the code completion task via learning from large-scale source code datasets. 
However, current approaches focus only on code context within the file or project, i.e. internal context. Our distinction is utilizing ``external'' context, inspired by human behaviors of copying from the related code snippets when writing code. Specifically, we propose a retrieval-augmented code completion framework, leveraging both lexical copying and referring to code with similar semantics by retrieval.
We adopt a stage-wise training approach that combines a source code retriever and an auto-regressive language model for programming language.
We evaluate our approach in the code completion task in Python and Java programming languages, achieving a state-of-the-art performance on CodeXGLUE benchmark.
\end{abstract}

\section{Introduction}




With the growth of software engineering field, large-scale source code corpus gives a chance to train language models in code domain \citep{hindle2016naturalness,tu2014localness}. And benefiting from the large transformer models \citep{vaswani2017attention} and pre-training techniques \citep{devlin2018bert,radford2018improving}, a rapid progress has been made in many code-related tasks like code search \citep{feng2020codebert,guo2020graphcodebert}, code summarization \citep{clement2020pymt5,ahmad2020transformer}, bug fixing \citep{mashhadi2021applying,drain2021generating} and code completion \citep{svyatkovskiy2020intellicode,liu2020multi,kim2021code,clement2020pymt5}. 

Code completion is considered as an essential feature towards efficient software development in modern Integrated Development Environments (IDEs). The task is formulated by predicting the following code token(s) based on the code context. 
Traditionally, code completion requires real-time program analysis and recommends type-correct code tokens \citep{tu2014localness}. 
Recently, statistical language models trained on large-scale source code data have shown high accuracy in the code completion task.
Primitive approaches take the given context only \citep{liu2016neural,karampatsis2020big}, some methods use richer information, e.g., adding code token types \citep{liu2020multi}, abstract syntax tree (AST) structures \citep{li2018code,kim2021code}, or extended hierarchical context \citep{clement2021long}.
However, one key limitation of existing methods is the scope of information they utilize; all the information is bounded in the given input file. This is unnatural from human perspective, as studies demonstrate that programmers tend to reuse an existing code snippet by copying part of code with or without minor modifications to accelerate software development \citep{roy2008empirical,baker2007finding}, leading a software repository usually containing 7-23\% cloned codes \citep{svajlenko2015evaluating}.

\begin{figure*}
    \centering
    \includegraphics[width=\textwidth]{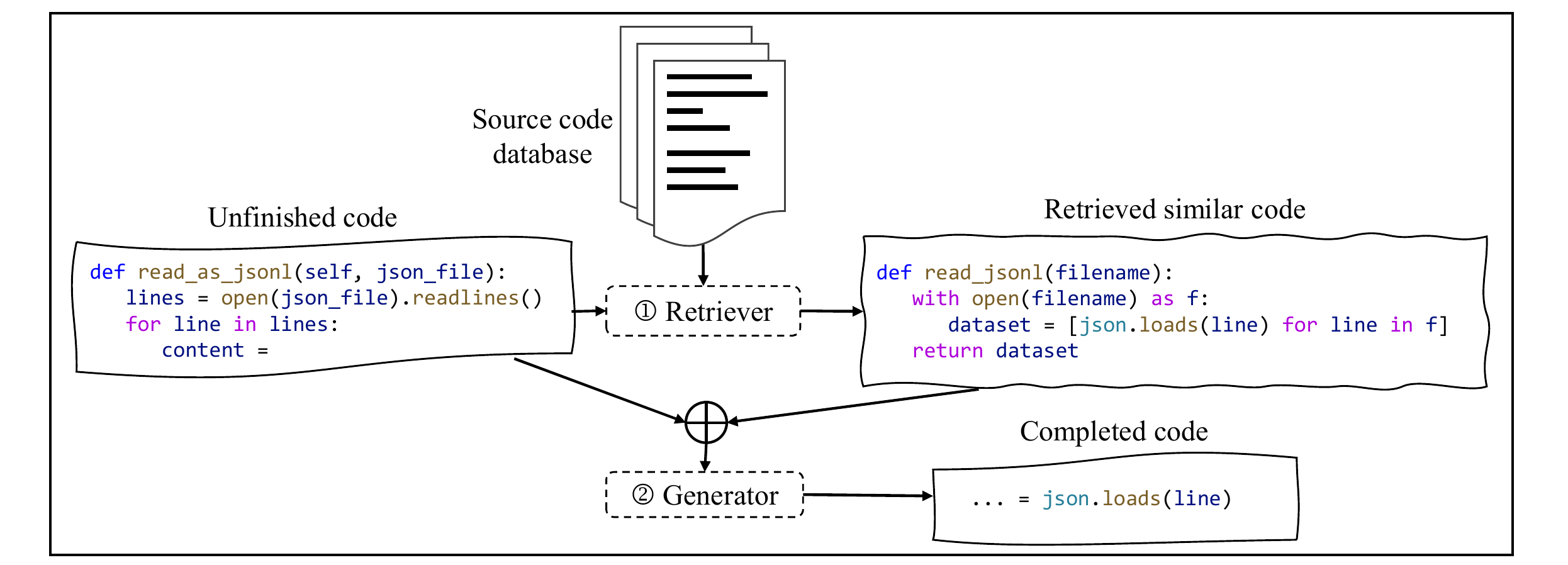}
    \caption{An illustration of ReACC framework. Given an unfinished code snippet to complete, ReACC first retrieves the similar code from source code database. Then the similar code is concatenated with the unfinished code, the completed code will be generated based on them.}
    \label{fig:illustration}
\end{figure*}

Motivated by this phenomenon, in this paper, we argue the utility of extending the information scope beyond the input file, i.e., into a large codebase. We conjecture that using codes with similar semantics as auxiliary information are beneficial to predict the following code tokens. Therefore, we propose ReACC -- a \textbf{Re}trieval-\textbf{A}ugmented \textbf{C}ode \textbf{C}ompletion framework (See Figure \ref{fig:illustration}). The code completion task under our framework can be re-formulated by, given a source code corpus for search and an unfinished code snippet to complete, using the unfinished code as a query to retrieve similar code snippets from search corpus, and predicting the following code tokens by reusing the retrieved code. 
ReACC consists of two core components: (1) a dual-encoder model served as the code-to-code search  \textbf{retriever} (2) an auto-regressive language model served as the code completion \textbf{generator}.
ReACC adopts the stage-wise training strategy which is widely used in other tasks like open-domain question answering \citep{karpukhin2020dense,izacard2021leveraging}, natural language to code generation \citep{hashimoto2018retrieve,parvez2021retrieval}, etc.


The simplest technique for retrieving code is to build a sparse vector retriever like TF-IDF or BM25 \citep{robertson2009probabilistic} which are both based on keyword matching algorithms. The sparse retriever can capture lexical information and is sensitive to the names of code identifiers. The dense retriever, on the contrary, can capture syntactic and semantic information by mapping a code snippet to a dense vector. In the code completion task, the code retriever is expected to comprehend the source code's intent in order to retrieve the semantically similar codes. On the other hand, considering programmers are prone to copy-and-paste existing code, the retriever should evaluate lexical similarity as well. To that end, we adopt the hybrid retriever \citep{karpukhin2020dense,ma2021replication}, which combines results of dense and sparse retriever. We employ a dual-encoder model architecture as the dense retriever since the cross-encoder model has a high computational complexity. 
To achieve a better understanding ability, we initialize our dense retriever with GraphCodeBERT \citep{guo2020graphcodebert}, which is a pre-trained BERT-based programming language understanding model. Then we continue pre-training the retriever by contrastive learning to enhance sentence embedding. As the labeled data containing similar code pairs is rare, we utilize various transformations to generate programs with similar functionality for data augmentation.

We implement the generator with a decoder-only transformer model. To incorporate the external information from retrieved similar code, we concatenate the obtained code and code context as input. The generator is initialized by CodeGPT-adapted \citep{lu2021codexglue} which is a domain-adaptation model from GPT-2 \citep{radford2018improving} pre-trained on code corpus.

We evaluate our ReACC framework on two benchmark datasets -- CodeXGLUE \citep{lu2021codexglue} and CodeNet \citep{puri2021project}, in Python and Java programming languages. ReACC achieves a state-of-the-art performance on both datasets. The experimental results demonstrate that external source code retrieved by our retriever is useful for auto-completing the partial code. 

To summarize, our main contributions are:
\begin{itemize}
    \item We propose a retrieval-augmented method to assist the code auto-completion task. \footnote{Our codes are available at \url{https://github.com/celbree/ReACC}}
    \item To adapt to the code completion scenario, where the retrieval query is an unfinished code snippet, we propose the partial code-to-code search task and create datasets for evaluation.
    \item We adopt semantic-preserving transformations for data augmentation to pre-train the code retrieval model.
\end{itemize}

\section{Related Work}

\subsection{Code completion}
Code completion is an essential task for code intelligence. 
\citet{hindle2016naturalness} are the first to use language model for code completion by N-gram technique. 
Deep neural networks \citep{liu2016neural,alon2020structural,karampatsis2020big} and pre-training approaches \citep{liu2020multi,svyatkovskiy2020intellicode} are later frequently utilized to accomplish this. 
Besides considering source code as code token sequences, some research focuses on completing an abstract syntax tree (AST) by anticipating the next node in the flattened tree \citep{li2018code,kim2021code}. 
\citet{guo2021learning} complete codes by generating sketches, i.e. code snippets with ``holes''.
\citet{svyatkovskiy2021fast} and \citet{clement2021long}, on the other hand, investigate ways to improve the efficiency and long-range modeling in the code completion task, respectively. 

All of these works employ previously written code context as inputs, along with AST structural information or token types. But none of them has attempted to leverage existing external code as auxiliary information.

\subsection{Retrieval on code intelligence}
\paragraph{Contrastive learning on code}
Inspired by the great success of contrastive learning in other domains \citep{wu2018unsupervised,reimers2019sentence,fang2020cert,chen2020simple,he2020momentum,radford2021learning,gao2021simcse}, researchers have deployed this technique to source code for better code fragment understanding. 
\citet{jain2020contrastive} and \citet{bui2021self} propose ContraCode and Corder, respectively. Both models use the self-supervised contrastive learning framework and generate code snippets as data augmentations via compiler-based semantic-preserving transformations. Their models have shown the effectiveness of contrastive learning in code clone detection, code search and code summarization tasks.
\textsc{SynCoBERT} \cite{wang2022syncobert} and UniXcoder \cite{guo2022unixcoder} are both pre-training models that utilize multi-modal data, including code, comment, and AST, for better code fragment representation through contrastive learning. 

\paragraph{Retrieval for code-related tasks}
Many code intelligence tasks benefit from information retrieval \citep{xia2017developers}. 
A common scenario for information retrieval in code domain is code search with natural language description as a query \citep{arwan2015source,gu2018deep,cambronero2019deep}.
As for other code intelligence tasks, 
\citet{hayati2018retrieval} propose an action subtree retrieval method called ReCode for
generating general-purpose code.
\citet{hashimoto2018retrieve} propose a retrieve-and-edit framework for code autocompletion and code generation. 
\citet{luan2019aroma} propose Aroma, which utilizes code-to-code structural search and intersecting candidate code snippets to recommend relevant code given another code snippet as a query.
Both \citet{wei2020retrieve} and \citet{li2021editsum} leverage the retrieve-and-edit/refine framework to improve model's performance in code summarization.
\citet{parvez2021retrieval} propose REDCODER, using a dense retriever trained on paired NL-code pairs to retrieve relevant comments or codes as a supplement for code summarization or code generation tasks. 

In most circumstances where a dense retriever is utilized, a natural language comment is treated as a query to retrieve code. In the code completion scenario, however, we focus on using code as query, particularly partial code, which is a more difficult task since there are few labeled data with semantically similar code pairs and in partial code search, semantics in query is incomplete.

\section{Approach}
We first introduce the formulation of retrieval-augmented code completion task. Then we give detailed descriptions on the retriever and generator in ReACC. We show how we continue pre-training GraphCodeBERT \citep{guo2020graphcodebert} with contrastive learning on code and how we address the problem that there is no labeled data for positive instances of similar programs in section \ref{sec:retrieval}. In section \ref{sec:generator} we talk about the way to aggregate retrieved code and code context in the generator.

\subsection{Task Formulation}\label{sec:task_def}
Assume that we have a source code database containing a large collection of software repositories, which consist of $D$ source code files, $f_1, f_2, ..., f_D$. Following the Dense Passage Retriever (DPR) model \citep{karpukhin2020dense}, we split each of the files into code fragments of equal lengths as the basic retrieval units. Such splitting not only leads to a better retrieval results as stated by \citet{karpukhin2020dense}, but also supports extreme long code files where each part of a file represents different semantics. Thus we get $M$ code fragments as the retrieval database $C = \{c_1, c_2, ..., c_M\}$. Let $X = \{x_1, x_2, ..., x_k\}$ be the unfinished code written previously, a retriever $\textbf{R}: (X,C) \rightarrow C$ retrieves the most similar code fragment $c_s$ in $C$. The generator $\textbf{G}$ predicts the following code token(s) $Y = \{x_{k+1}, ..., x_{k+n}\}$, where $n=1$ in the token-level code completion task, based on context and retrieved code. Formally, $P(Y)=\prod_{i=1}^{n} P(x_{k+i}|c_s,x_{1:k+i-1})$.

\begin{figure}
    \centering
    \includegraphics[width=0.48\textwidth]{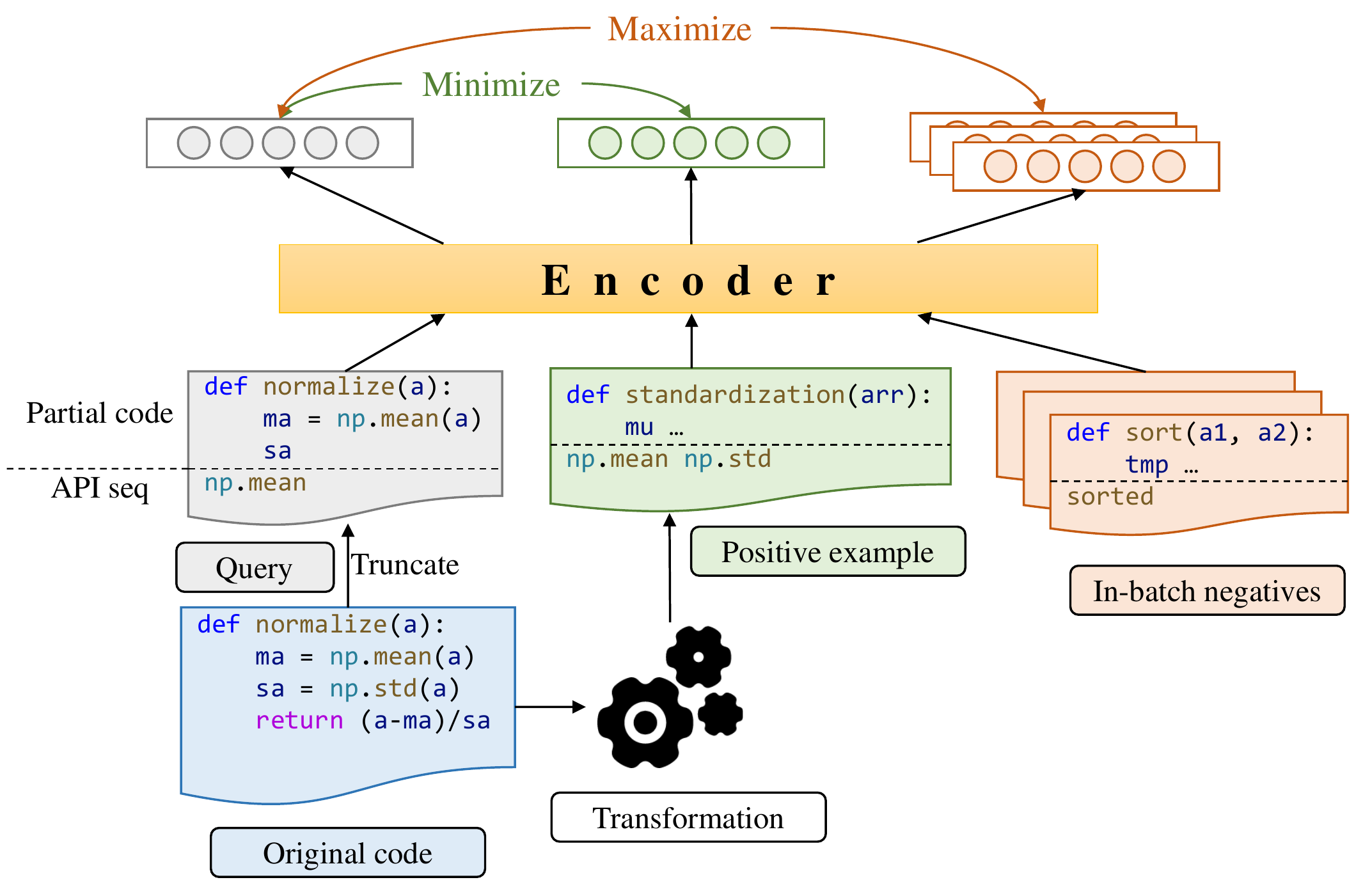}
    \caption{Illustration on the training process of the retriever in our proposed framework ReACC. }
    \label{fig:retriever}
\end{figure}

\begin{figure*}[htbp]
    \centering
    \includegraphics[width=\textwidth]{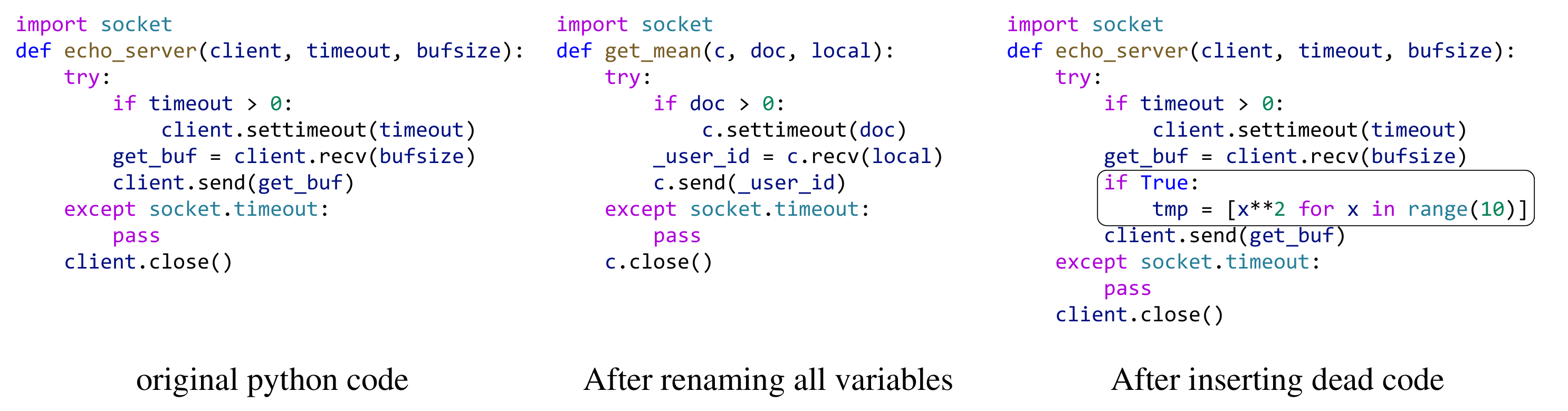}
    \caption{An example of applying semantic-preserving transformations to Python code. }
    \label{fig:trans_eg}
\end{figure*}

\subsection{Retriever}\label{sec:retrieval}
The retrieval module in ReACC is expected to retrieve semantically equivalent code given an incomplete code. 
We adopt the hybrid retriever \citep{karpukhin2020dense,ma2021replication} framework by combining scores of sparse and dense retriever. The sparse retriever we use is BM25 \citep{robertson2009probabilistic} based on the implementation of ElasticSearch\footnote{https://github.com/elastic/elasticsearch}. As a term-based retrieval method, BM25 considers each code fragment as a code token sequence and employs bag-of-words representations. The matching score computed by BM25 indicts lexical similarity between the query and document. As for the dense retriever, it maps each code fragment to a $d$-dimension dense vector. We construct it in this paper based on the DPR model \citep{karpukhin2020dense}. Figure \ref{fig:retriever} illustrates the training process of the dense retriever of ReACC. In the following, we will walk through it in detail.


\paragraph{Dense Retriever} 
Our dense retriever consists of two bidirectional transformer-based encoders $E_C$ and $E_Q$. $E_C$ encodes each code fragment in the retrieval database $C$ and builds indexes for them. The query is encoded by $E_Q$. We take the representation of \texttt{[CLS]} token as output and the similarity is computed by $sim(q, c) = E_C(c)^TE_Q(q)$.
Since both $E_C$ and $E_Q$ take source code as inputs with the only difference being whether they are partial or not, the dual encoders share weights in ReACC.
At the training stage, following DPR \citep{karpukhin2020dense}, we adopt in-batch negatives to calculate the contrastive loss by InfoNCE \citep{oord2018representation}:
\begin{equation}
\begin{aligned}
&L(q, c^+, c^-_1, c^-_2, ..., c^-_m) \\ =
&-log\frac{e^{sim(q, c^+)}}{e^{sim(q, c^+)} + \sum^m_{i=1}e^{sim(q, c^-_i)}}
\end{aligned}
\end{equation}
However, unlike DPR, we don't employ "hard" negatives which are retrieved from BM25. Because programmers tend to copy tokens directly, a code with distinct semantics but substantial lexical similarity can help with code completion.

\paragraph{Data Augmentation}
The purpose of contrastive learning of the dense retriever in ReACC is to learn a representation of code fragments that keeps codes with similar or equivalent semantics close and dissimilar codes far apart. It requires numerous positive and negative code pairs.
However, it is difficult to identify similar programs based on an unlabeled code corpus, e.g., certain widely used datasets \citep{allamanis2013mining,raychev2016probabilistic,husain2019codesearchnet} mined from GitHub repositories. 

Searching semantically equivalent code requires extra code compilation and execution costs \citep{massalin1987superoptimizer,churchill2019semantic}, which is unrealistic in a large database. Instead of searching, an alternative way is to create code snippets with same functionalities for data augmentation. To do so, we apply several semantic-preserving transformations to the original source code to construct a set of variants. There exists several attempts to apply such transformation to code \citep{jain2020contrastive,rabin2021generalizability,bui2021self}. In this paper, we mainly adopt identifier renaming and dead code (unreachable or unused code) insertion. Figure \ref{fig:trans_eg} shows an example of performing such transformations to a Python code.
\begin{itemize}[leftmargin=*]
    \item \textbf{Identifier renaming} is a method of renaming an identifier with another. 
    We only rename variable and method names as other identifiers cannot be changed arbitrarily like built-in types or API calls.
    Different from previous works, we preserve part of the lexical information while modifying the names at the same time based on the consideration that identifier names typically convey the meanings for humans and lexical similarity contributes a lot for retrieving (It is verified in section \ref{sec:clone_res}). To do so, we mask all the identifiers in a program and leverage GraphCodeBERT \citep{guo2020graphcodebert} to predict each identifier like in the masked language model task. The top-10 predictions (excluding the original identifier) are selected as the candidate set for renaming.
    \item \textbf{Dead code insertion} is to insert a dead code into a code fragment at a proper location. 
    Dead code is a code snippet which can never be reached \citep{xi1999dead} or is reachable but whose result can never be used in any other computation \citep{debray2000compiler}. 
    In software engineering, dead code insertion is one of the most common techniques for code obfuscation \citep{you2010malware}, whose goal is to modify a code to make it hard to understand but remain its functionality, which is similar to our goal.
    We first randomly select variable names which don't appear in this program and then use them to form a statement from a predefined set of dead code (See Appendix \ref{app:dead} for details), such as assignment, method invocations, looping statement, conditional statement and so on. We traverse the AST and identify all the statements. Then we choose a statement at random and insert the dead code after it, leading a new subtree in the AST.
\end{itemize}


\paragraph{Input Format} We integrate both the code token sequence and the API usage sequence as inputs. API usage sequence is highly related to the functionality of a code snippet \cite{gu2016deep,hu2018summarizing}. To improve the code representation, we extract the API sequence and append it to the source code token sequence. 
Finally, we use a random truncation of the original code as the query and the entire created program as the positive example during training to address the problem on how to retrieve based on incomplete semantics.


\subsection{Generator}\label{sec:generator}
The output of retriever is the retrieved code $c_s$. Considering $c_s$ is queried by code context $x$ while our target is the following code of $x$, so we propose \textit{fragment alignment} -- using the next fragment $c_s'$ of $c_s$ in the same file (we have split each file into code fragments for retrieval as discussed in Section \ref{sec:task_def}) for completing the next fragment of $x$. Thus, the input sequence for the generator is the concatenation of $c_s'$ and  $x$: $x' = c_s' \oplus x$.

The generator module in ReACC supports any model architecture that can perform code completion task. In our experiments, we adopt CodeGPT-adapted \citep{lu2021codexglue}, which is a decoder-only transformer model pre-trained on Python and Java datasets from CodeSearchNet \citep{husain2019codesearchnet} via casual language model. CodeGPT-adapted has shown promising results in the code completion task in CodeXGLUE benchmark \citep{lu2021codexglue} on two widely used code completion datasets.


\section{Experiments: Code Clone Detection}
In order to evaluate the effectiveness of the code-to-code retrieval module in ReACC, we perform code clone detection task which aims to retrieve semantic equivalent programs. In this section, we describe how we create the test dataset for this task and how we evaluate the performance of ReACC's retriever.

\begin{table*}
    \centering
    \begin{small}
    \begin{tabular}{lcccccc}
         \hline
         {\textbf{Dataset}} & \textbf{Language} & \textbf{Task} & \textbf{Train} & \textbf{Valid} & \textbf{Test} & \textbf{Desc.} \\
         \hline
         \multirow{3}{*}{\makecell[l]{CodeNet \\ \citep{puri2021project}}} & Python & Clone & - & - & 15,594 & Solutions for 2,072 problems \\
         & Java & Clone & - & - & 14,426 & Solutions for 1,599 problems \\
         & Python & Completion & 2,636,118 & 32,984 & 10,000 & For line-level completion \\
         \hline
         \multirow{3}{*}{\makecell[l]{CodeXGLUE \\ \citep{lu2021codexglue}}} & Python & Completion & 95,000 & 5,000 & 50,000 / 10,000 & Use PY150 \\
         & Python$^\dagger$ & Completion & 95,000 & 5,000 & - / 20,000 & Applying eWASH \\
         & Java & Completion & 12,934 & 7,176 & 8,268 / 3,000 & Use JavaCorpus \\
         \hline
    \end{tabular}
    \end{small}
    \caption{Dataset statistics. The two numbers in \textbf{Test} of CodeXGLUE denote the examples for token-level and line-level code completion, respectively. $^\dagger$ is a newly created test set, see the text for details.}
    \label{tab:dataset}
\end{table*}

\subsection{Dataset}
\paragraph{CodeNet}\citep{puri2021project} dataset consists of a large collection of programs which are derived from online judge websites. We respectively create a code clone detection evaluation dataset from CodeNet in Python and Java with zero-shot setting. We collect code solutions for thousands problems and solutions for the same problem are considered as semantically equivalence. The data statistics are shown in Table \ref{tab:dataset}.

\paragraph{Retrieval Training Set}
The dense retriever in ReACC is pre-trained on CodeSearchNet dataset \citep{husain2019codesearchnet}, a large-scale source code corpus extracted from GitHub repositories. We employ 1.6M Java methods and 1.2M Python functions from it.

\subsection{Baseline Methods}
\paragraph{CodeBERT}\citep{feng2020codebert} is a pre-trained model for programming language, which is trained on NL-PL pairs from CodeSearchNet dataset in six programming languages. 

\paragraph{GraphCodeBERT}\citep{guo2020graphcodebert} is also pre-trained on CodeSearchNet NL-PL pairs and considers the inherent structure of code i.e. data flow. 

\subsection{Experiment Setup}
The retrieval encoder is initialized with GraphCodeBERT. It is continual pre-trained with both masked language model objective and contrastive learning. We use in-batch negatives with a batch size of 256. With a learning rate of 5e-5, We train the retriever for Python and Java for 30 epochs each.

We implement the code clone detection experiment in the \textit{partial search} way, 
which is ideally adapted to code completion scenarios as it accepts a partial program as a query while maintaining the same goal.


\begin{table*}
    \centering
    \begin{small}
    \begin{tabular}{ccccc}
         \hline
         \multirow{2}{*}{\textbf{Model}} & \multicolumn{2}{c}{Python} & 
         \multicolumn{2}{c}{Java} \\
         \cline{2-5}
         & MAP@100 & Precision & MAP@100 & Precision \\
         \hline
         CodeBERT & 1.47 & 4.75 & 1.15 & 4.58 \\
         GraphCodeBERT & 5.31 & 15.68 & 4.54 & 16.05 \\
         BM25 & \textbf{10.32} & 23.17 & 8.67 & 25.85 \\
         ReACC-retriever & 9.60 & \textbf{27.04} & \textbf{9.31} & \textbf{27.55} \\
         \hline
    \end{tabular}    
    \end{small}
    \caption{Results on zero-shot code clone detection dataset created from CodeNet.}
    \label{tab:clone}
\end{table*}

\subsection{Results} \label{sec:clone_res}
Table \ref{tab:clone} shows the results in the zero-shot code clone detection task on CodeNet dataset, with the \textit{partial search} setting. Models are measured by MAP@K (Mean Average Precision at K), which is the evaluation metric in the CodeXGLUE clone detection task, and precision at 1, as we only care about the most similar code for code completion. 
From the comparison with other transformer-based encoders, we can see
CodeBERT and GraphCodeBERT 
can hardly retrieve equivalent code. While our model significantly outperforms them, which indicts our model is capable of retrieving the semantically equivalent code even when the  query's semantics is incomplete.

We also find that BM25 performs splendidly in this task,
which is quite different from the performance on other tasks like open-domian QA \citep{karpukhin2020dense}, code summarization \citep{parvez2021retrieval}, etc. 
The findings suggest that semantically related codes are likely to be lexically similar, which leads lexical similar to contribute more for retrieval, making code-to-code search easier than text-to-code or question-to-passage search using the term-based retrieval method.


\section{Experiments: Code Completion}
In this section, we evaluate ReACC on end-to-end code completion.

\begin{table*}
    \centering
    \begin{small}
    \begin{tabular}{ccccccc}
         \hline
         \multirow{2}{*}{\textbf{Model}} & \multicolumn{3}{c}{PY150} & 
         \multicolumn{3}{c}{JavaCorpus} \\
         \cline{2-7}
         & Perplexity & Exact Match & Edit Sim & Perplexity & Exact Match & Edit Sim \\
         \hline
         GPT-2 & - & 41.73 & 70.60 & - & 27.50 & 60.36 \\
         CodeGPT & 2.502 & 42.18 & 71.23 & 4.135 & 28.23 & 61.81 \\
         CodeGPT-adapted & 2.404 & 42.37 & 71.59 & 3.369 & 30.60 & 63.45 \\
         CodeT5-base & - & 36.97 & 67.12 & - & 24.80 & 58.31 \\
         PLBART & - & 38.01 & 68.46 & - & 26.97 & 61.59 \\
         \hline
         ReACC-bm25 & 2.312 & 46.07 & 73.84 & 3.352 & 30.63 & 64.28 \\
         ReACC-dense & 2.329 & 45.32 & 73.95 & 3.355 & 30.30 & 64.43 \\
         ReACC-hybrid & \textbf{2.311} & \textbf{46.26} & \textbf{74.41} & \textbf{3.327} & \textbf{30.70} & \textbf{64.73} \\
         \hline
    \end{tabular}    
    \end{small}
    \caption{Results on the code completion task in CodeXGLUE}
    \label{tab:codexglue}
\end{table*}

\begin{table}
    \centering
    \begin{small}
    \begin{tabular}{ccc}
         \hline
         & \textbf{Exact Match} & \textbf{Edit Sim} \\
         \hline
         GPT-2 & 37.08 & 68.71 \\
         CodeGPT & 37.21 & 69.00 \\
         CodeGPT-adapted & 38.77 & 70.07 \\
         X-CodeGPT & 39.41 & 70.97 \\
         \hline
         ReACC-bm25 & \textbf{40.24} & 71.65 \\
         ReACC-dense & 39.67 & 71.80 \\
         ReACC-hybrid & 40.15 & \textbf{72.01} \\
         \hline
    \end{tabular}    
    \end{small}
    \caption{Results on the new testset created from PY150 in CodeXGLUE}
    \label{tab:newpy}
\end{table}

\subsection{Dataset}
\paragraph{CodeXGLUE}\cite{lu2021codexglue} is a benchmark dataset containing 14 datasets for 10 diversified code intelligence tasks. We use PY150 dataset \citep{raychev2016probabilistic} in Python and GitHub Java Corpus dataset \citep{allamanis2013mining} in Java from it for code completion task. Table \ref{tab:dataset} shows the data statistics.

\subsection{Baseline Methods}
\paragraph{CodeGPT/CodeGPT-adapted}\citep{lu2021codexglue} are both pre-trained on Python and Java datasets from CodeSearchNet. CodeGPT is trained from scratch while CodeGPT-adapted is a domain adaptation model which is initialized by GPT-2 \citep{radford2019language}.

\paragraph{PLBART}\citep{ahmad2021unified} is based on BART \citep{lewis2020bart} architecture which employs denoising sequence-to-sequence (Seq2Seq) pre-training and is pre-trained on unlabeled data across PL and NL.

\paragraph{CodeT5}\citep{wang2021codet5} is also an encoder-decoder pre-trained model which adapts T5 \citep{raffel2019exploring} architecture and considers the identifier-aware token type information in code.

\paragraph{X-CodeGPT} is a variant of CodeGPT which adapts eWASH \citep{clement2021long} to CodeGPT. \citet{clement2021long} propose eWASH, a method for leveraging the syntax hierarchy of source code to give the model wider field of vision in a file and achieving a new SOTA performance on the CodeXGLUE code completion task. We reproduce their method and develop X-CodeGPT by adapting eWASH to CodeGPT-adapted.  

\subsection{Experiment Setup}
\paragraph{Fine-tune} 
We fine-tune CodeGPT-adapted on PY150 and GitHub Java Corpus datasets, respectively, and use it as the generator in ReACC. The number of epochs for training PY150 is 30 and Java Corpus is 10, with a batch size of 96 and a learning rate of 2e-5. Except for X-CodeGPT, all other baseline models are fine-tuned with the same settings.

As for X-CodeGPT, we pre-train it with a training set extracted from CodeSearchNet in eWASH format, where each example is a function body with its corresponding extended context, as described by \citet{clement2021long}. Since eWASH requires codes parsed into ASTs but codes in CodeXGLUE have been tokenized and cannot be parsed, we build a new dataset from PY150 to fine-tune X-CodeGPT on CodeXGLUE. As a result, we download the origin files in PY150 and create a new dataset that retains the train/valid/test split, as seen in Table \ref{tab:dataset}.

\paragraph{Evaluation}
Following \citet{lu2021codexglue}, we conduct two code completion scenarios, token-level and line-level completion, to measure models' ability of predicting one and more tokens. Perplexity is the evaluation metric for token-level completion, whereas exact match accuracy (EM) and edit similarity are used for line-level completion. For token-level completion, based on the consideration of efficiency, instead of applying retrieval at each step, we retrieve similar codes based on current context after predicting the first 100 tokens, and leverage it for further prediction.

\paragraph{Retrieval Database}
We use the training set of PY150 and Java Corpus as retrieval database for test. We don't use the contrastive pre-training corpus (i.e., CodeSearchNet) in order to avoid the duplication between CodeXGLUE and CodeSearchNet as they are both extracted from GitHub.

\paragraph{Hybrid Retriever}
A linear combination of scores from BM25 and our dense retriever forms a hybrid retriever. Specifically, we calculate the score by $sim(q,c)+\alpha \cdot BM25(q,c)$ and let $\alpha = 0.9$ based on the results on dev set for both PY150 and Java Corpus datasets.

\subsection{Results}
Table \ref{tab:codexglue} and Table \ref{tab:newpy} compare different baseline models on code completion task in the CodeXGLUE Python and Java datasets. ReACC framework with the hybrid retriever outperforms consistently than other baselines on all datasets, which proves our conjection that the ``external'' context is beneficial to the code completion task. The comparison with X-CodeGPT in Table \ref{tab:newpy} demonstrates that utilizing ``external'' context could be more useful than making the most of the current code file. Among three configurations of the retriever in ReACC, hybrid retriever performs best on almost all metrics except the exact match score in the new test set of PY150. 

From Table \ref{tab:codexglue}, we can observe that comparing the two datasets, the improvement in the PY150 dataset is greater than that in the Java Corpus dataset. The reason for this is that the retrieval database for Java (i.e., the training set) is much smaller. The CodeXGLUE Java Corpus dataset contains only 12,934 files for training so that it's more difficult to retrieve similar code from them.

Another finding is that BM25 shows comparable results with dense retriever and even performs better in perplexity and exact match metrics. The findings indict that the code completion task can benefit from both semantically and lexically similar codes.

\begin{table}
    \centering
    \begin{small}
    \begin{tabular}{ccc}
         \hline
         & \textbf{Exact Match} & \textbf{Edit Sim} \\
         \hline
         CodeGPT-adapted & 46.38 & 74.10 \\
         \hline
         ReACC-bm25 & 55.88 & 79.62 \\
         ReACC-dense & 64.21 & 84.57 \\
         ReACC-hybrid & \textbf{64.74} & \textbf{84.93} \\
         \hline
    \end{tabular}    
    \end{small}
    \caption{Results on the code completion task created from CodeNet Python dataset}
    \label{tab:codenet}
\end{table}

\begin{table}
    \centering
    \begin{small}
    \begin{tabular}{lcc}
         \hline
         & \textbf{EM} & \textbf{Edit Sim} \\
         \hline
         ReACC-dense & \textbf{45.32} & \textbf{73.95} \\
         \hline
         Retriever & & \\
         {- identifier renaming} & 44.91 & 73.14 \\
         {- dead code insertion} & 45.11  & 73.57 \\
         {- API sequence} & 44.77 & 73.01 \\
         {- query truncation} & 43.93 & 72.65 \\
         Generator & & \\
         {- fragment alignment} & 45.08 & 73.56 \\
         \hline
    \end{tabular}    
    \end{small}
    \caption{Ablation study for both retriever and generator module. Experiments are run in CodeXGLUE PY150 dataset.}
    \label{tab:ablation}
\end{table}

\subsection{Analysis}
\paragraph{ReACC in specific domain}
Both PY150 and Java Corpus datasets are extracted from GitHub repositories which are distributed in a wide domain. As some people frequently write codes in a more specific domain, e.g., data mining/pattern recognition domain for Kaggle\footnote{https://www.kaggle.com/} users, algorithm domain for ACM community, etc. To evaluate ReACC in a specific code domain, we construct a code completion Python dataset from CodeNet, which can be considered in algorithm domain. Table \ref{tab:codenet} reveals that ReACC significantly outperforms CodeGPT-adapted in CodeNet by 10\% and 18\% absolute improvement in edit similarity and exact match, respectively. According to the findings, ReACC is more effective in a specific domain. We also notice that ReACC with dense retriever outperforms BM25 significantly in CodeNet. It can be explained by the fact that in algorithm domain, semantically similar code may be more valuable than for code completion lexically similar code.

\paragraph{Ablation study}
To further understand how our training options affect model performance, we conduct ablation experiments. As seen in Table \ref{tab:ablation}, when data argumentation and training strategies in retriever or generator are eliminated, the metrics degrade. The most essential factor among them is query truncation. Comparing the two semantic-preserving transformations, identifier renaming contributes more than dead code insertion.When fragment alignment is removed from generator, i.e. using the retrieved code snippet itself for generator, performance suffers slightly.

\paragraph{ReACC vs GitHub Copilot}
GitHub Copilot\footnote{https://copilot.github.com/} is a powerful technique for code completion which uses OpenAI Codex \citep{chen2021evaluating} as the model backend. 
We run some qualitative examples with its extension in VSCode, which are shown in the Appendix \ref{app:examples}. 
It worth noting that Codex is more powerful than CodeGPT since it is a large-scale pre-trained model that is trained on all source codes in GitHub based on GPT-3 \citep{brown2020language}. 
However, in some cases, ReACC with CodeGPT as the generator outperforms Copilot. And in \ref{fig:ex2} Copilot itself can benefit from ReACC when it takes advantage of ReACC's retriever, which indicates the effectiveness of retrieval-augmented method for strong generative models.

\section{Conclusion}
We propose ReACC, a retrieval-augmented code completion framework that utilizes ``external'' context for the code completion task by retrieving semantically and lexically similar codes from existing codebase. We pre-train a dual-encoder as a retriever for partial code search, which retrieves code fragments given a partial code. Our method can adopt any architecture that can perform code completion as the generator. On the CodeXGLUE benchmark, ReACC achieves a state-of-the-art performance in the code completion task.

\section*{Acknowledgements}
This work is supported by Microsoft Research Asia and IITP grants (2021-0-01696, High Potential Individuals Global Training Program)

\bibliography{custom}
\bibliographystyle{acl_natbib}

\appendix

\section{Predefined Dead Code}\label{app:dead}
\begin{figure*}[htbp]
    \centering
    \includegraphics[width=\textwidth]{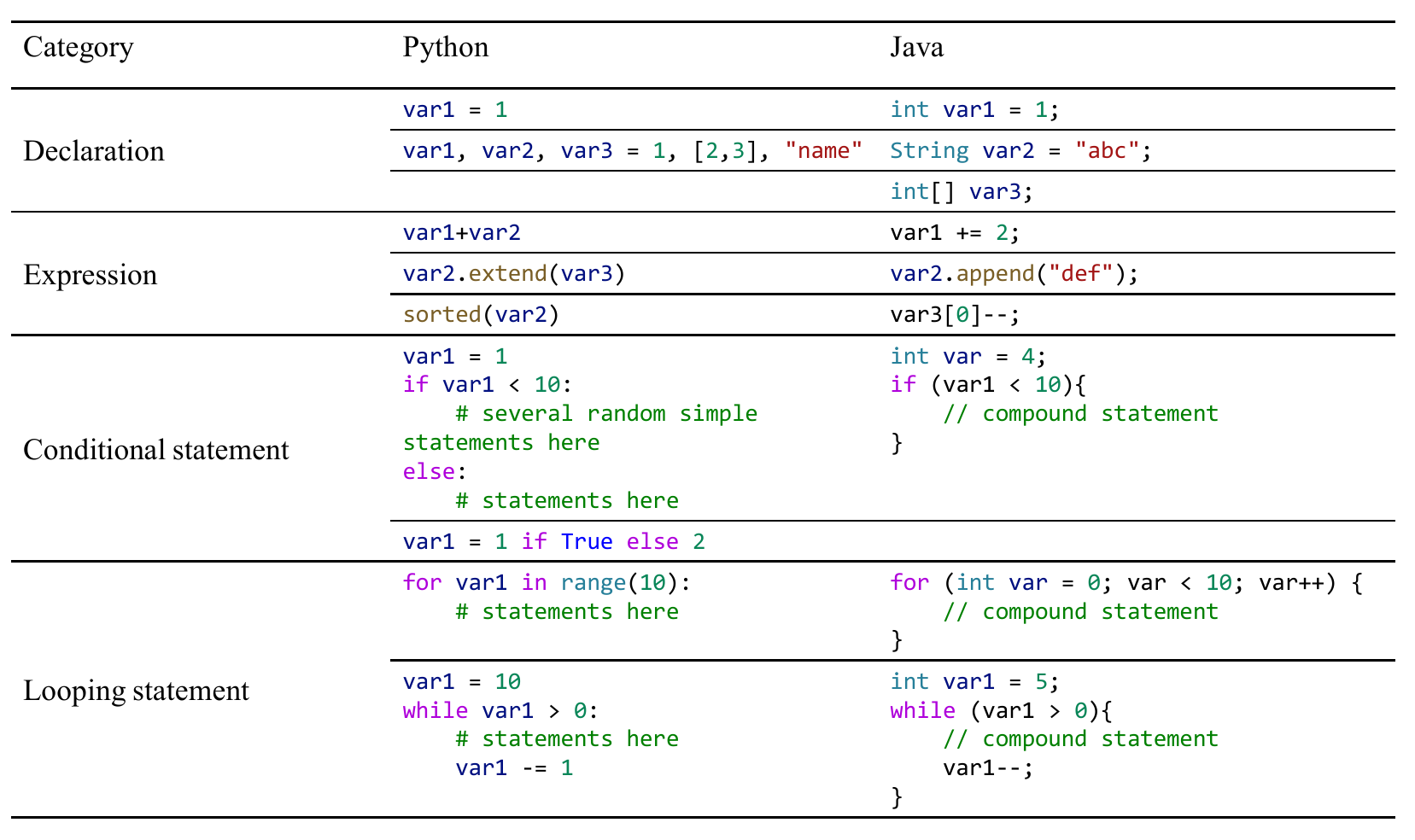}
    \caption{Examples of predefined set of dead code. Vars are randomly selected from other files. Literals like strings and integers are also generated at random. For conditional and looping statements, several simple statements (i.e., declaration and expression) are generated to fill the body.}
    \label{fig:deadcode}
\end{figure*}

We define a set of dead code to choose from for both Python and Java. We focus on four kinds of common statements, i.e., declaration statement, expression statement, conditional statement and looping statement. Examples are shown in figure \ref{fig:deadcode}. To generate a dead code snippet, we can use one kind of them or combine different statements together. 

\section{Qualitative Examples}\label{app:examples}

\begin{figure*}[htbp]
    \centering
    \includegraphics[width=\textwidth]{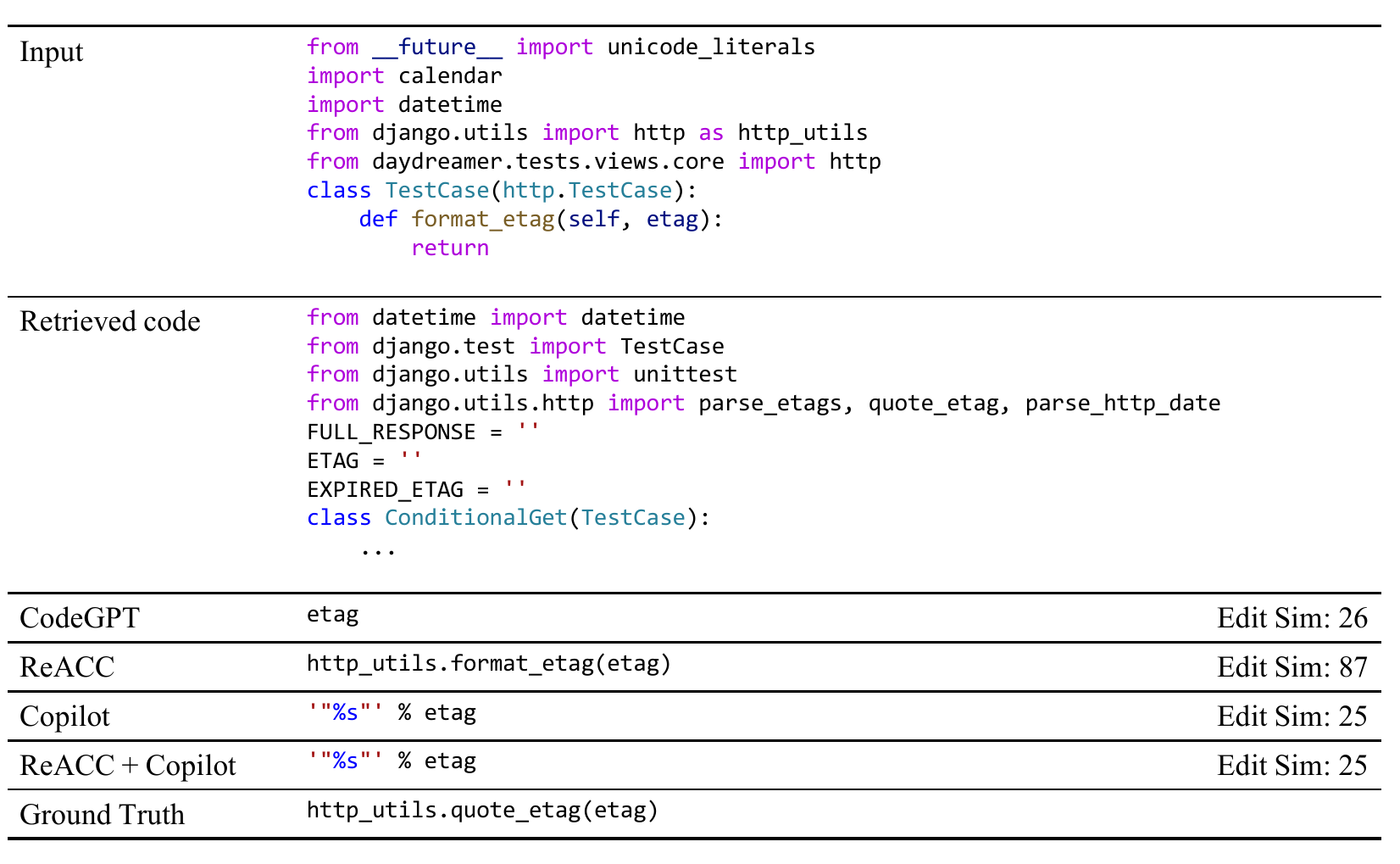}
    \caption{An qualitative example from PY150 test set. The input code comes from \url{https://github.com/skibblenybbles/django-daydreamer/blob/master/daydreamer/tests/views/behaviors/http/base.py}}
    \label{fig:ex1}
\end{figure*}

\begin{figure*}[htbp]
    \centering
    \includegraphics[width=\textwidth]{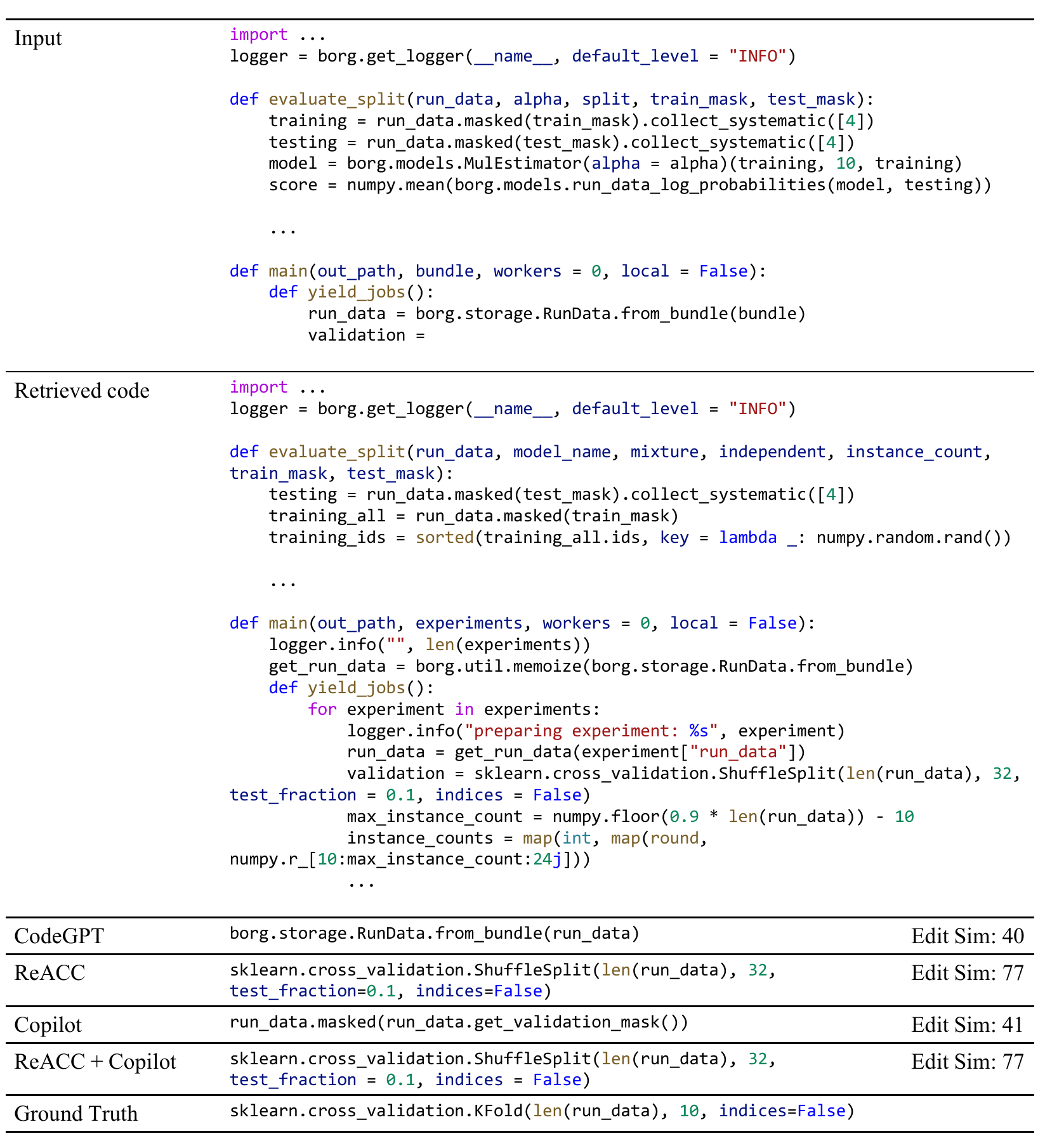}
    \caption{An qualitative example from PY150 test set. The input code comes from \url{https://github.com/borg-project/borg/blob/master/borg/experiments/mul_over_alpha.py}}
    \label{fig:ex2}
\end{figure*}

Figure \ref{fig:ex1} and figure \ref{fig:ex2} show qualitative examples of generated code by different models. ReACC + Copilot denotes ReACC framework with Copilot as the generator.

\end{document}